# Flexible delivery of high-power picosecond laser in purely-single optical mode of anti-resonant hollow-core fiber for micromachining

Xinshuo Chang,[1,2,3,4,†] Qinan Jiang,[1,†] Zhiyuan Huang,[1,2,*] Jinyu Pan,[2] Qingwei Zhang,[5,6] Nan Li,[5] Zhuozhao Luo,[1] Ruochen Yin,[1,3] Wenbin He,[1] Jiapeng Huang,[1] Yuxin Leng,[2,4] Xin Jiang,[1] Shanglu Yang,[5] and Meng Pang[1,2,7]

[1]Russell Centre for Advanced Lightwave Science, Shanghai Institute of Optics and Fine Mechanics (SIOM) and Hangzhou Institute of Optics and Fine Mechanics (HIOM), Hangzhou 311421, China
[2]State Key Laboratory of High Field Laser Physics and CAS Center for Excellence in Ultra-intense Laser Science, Shanghai Institute of Optics and Fine Mechanics (SIOM), Chinese Academy of Sciences (CAS), Shanghai 201800, China
[3]Center of Materials Science and Optoelectronics Engineering, University of Chinese Academy of Sciences, Beijing 100049, China
[4]Hangzhou Institute for Advanced Study, Chinese Academy of Sciences, Hangzhou 310024, China
[5]Center for Laser-Aided Manufacturing and Processing, Shanghai Institute of Optics and Fine Mechanics (SIOM), Chinese Academy of Sciences (CAS), Shanghai 201800, China
[6]e-mail: zhangqingwei@siom.ac.cn
[7]e-mail: pangmeng@siom.ac.cn
*Corresponding author: huangzhiyuan@siom.ac.cn
†These authors contributed equally to this work.

**We present the flexible delivery of picosecond laser pulses with up to 20 W average power over a 3-m-long sample of anti-resonant hollow-core fiber (AR-HCF) for laser micromachining applications. Our experiments highlight the importance of optical mode purity of the AR-HCF for the manufacturing precision. We demonstrate that compared with an AR-HCF sample with a capillary to core (d/D) ratio of ~0.5, the AR-HCF with a d/D ratio of ~0.68 exhibits better capability of high-order-mode suppression, giving rise to improved micromachining quality. Moreover, the AR-HCF delivery system exhibits better pointing stability and set-up flexibility than the free-space beam delivery system. These results pave the way to practical applications of AR-HCF in developing advanced equipment for ultrafast laser micromachining.**

Ultrafast lasers, capable of generating optical pulses with ultrashort pulse widths and high peak powers, play crucial roles in the field of precision manufacturing, such as cutting [1], welding [2] and surface treatment [3]. Although high-reflection mirrors and retroreflective optics can be used to achieve high-efficiency delivery of ultrafast pulses, such free-space systems exhibit, generically, high sensitivity to environments and require regular maintenances. Additionally, the relatively complex set-up somewhat limits the flexibility and stability of free-space laser beam delivery systems, restricting the application scope of advanced laser-micromachining equipment. Consequently, flexible delivery of high-power ultrafast laser over optical fiber is highly required for laser micromachining applications [4-6].

Even though conventional solid-core fiber, as a mature technique for high-average-power laser delivery [7,8], has been widely used for transmission of continuous-wave (CW) or quasi-CW laser light, high-fidelity transmission of high-power ultrafast pulses with μJ-level pulse energies over such solid-core fiber is impossible. Due to the extremely-high peak power on the fiber-end surface and excessive nonlinearity accumulation over the fiber length, disastrous material damage and pulse distortion cannot be avoided during ultrafast pulse propagation in the solid-core fiber. In contrast, the development of anti-resonant hollow-core fiber (AR-HCF) [9-12] provides a promising possibility of high-fidelity delivery of high-power ultrafast laser pulses, thanks to its intrinsic features of ultra-low nonlinearity, ultra-high damage threshold and weak waveguide dispersion [13-15]. Although a few studies on high-peak-power laser pulse delivery over AR-HCF, have been demonstrated in recent several years [16-18], the performance of such an AR-HCF delivery system in practical laser micromachining, especially the influence of transmitted optical-mode purity on the manufacturing quality, has not been studied yet.

In this Letter, we studied the AR-HCF delivery of 20 W picosecond laser pulses at 1064 nm in laser micromachining experiments, illustrating, for the first time to our knowledge, the importance of the fiber optical-mode quality for micromachining quality. We demonstrated experimentally that the AR-HCF with a capillary to core (d/D) ratio of ~0.68 [19] can efficiently overcome the problem of high-order-mode excitation due to the movement and sway of the fiber sample. We found that the ultrafast laser light, after flexible delivery over d/D=~0.68 AR-HCF sample, exhibit better optical-mode purity under external perturbations than that delivered over d/D=~0.5 AR-HCF sample. This improved mode stability results in perfect circularity of the manufactured spots with better consistency in the single-shot micromachining experiment, and in the in-line processing experiment, the good mode purity leads to relatively straight processing lines, similar with those achieved using a free-space delivery system. Our findings highlight the importance of high optical-mode purity of the AR-HCF for improving the laser manufacturing quality.

of the fiber after encasing is ~30 cm. At the output end of the AR-HCF inside the micromachining unit (see Fig. 1), the laser beam was collimated by a plano-convex lens (L2) with a focal length of 7.5 cm. Through the moving-mirror system (M4 – M6), the laser beam was then focused onto the surface of the material under processing (thick aluminum sheets in our experiments) by an F-theta lens. Two AR-HCF samples with different parameters (AR-HCF1: d = 16.6 μm, D = 33 μm, d/D = ~0.5 and AR-HCF2 with improved optical-mode purity [19,20]: d = 19.9 μm, D = 29.2 μm, d/D = ~0.68), were used in the experiment, and the scanning electron microscope (SEM) images of the two AR-HCF samples are shown as the inset of Fig. 1.

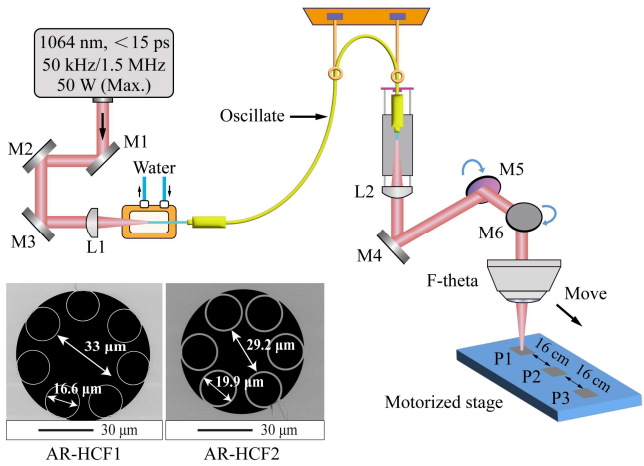

Fig. 1. Experimental set-up. M1-M4 are optical reflectors with a reflectivity of >99.5%; M5 and M6 are moving mirrors; L1 and L2 are plano-convex lenses with focal lengths of 7.5 cm. The AR-HCF is encased within a protective metallic cable, and the input end of the AR-HCF is installed on a copper plate. In the micromachining unit, three aluminum sheets are placed on the processing table. Inset: SEM images of AR-HCF1 and AR-HCF2, with d/D values of ~0.5 and ~0.68, respectively.

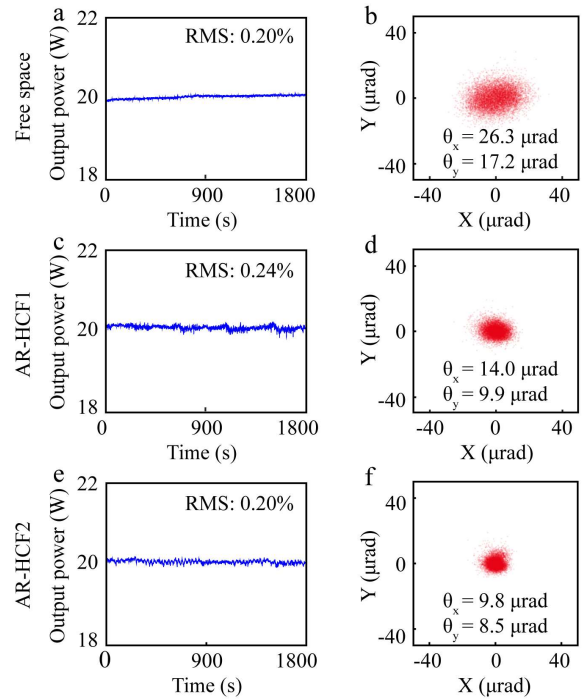

Fig. 2. The power stability (a) and pointing stability (b) of free-space laser delivery system, respectively; (c), (d) Results of the 3-m-long AR-HCF1 laser delivery system; (e), (f) Results of the 3-m-long AR-HCF2 laser delivery system.

The experimental set-up is sketched in Fig. 1. The light source was a commercial picosecond solid-state laser (HyperRapid NX, 1064-50) with a lasing wavelength of 1064 nm, a tunable repetition rate up to 1.5 MHz, a pulse width of <15 ps and a spectral bandwidth of <0.11 nm. The maximum output average power of the laser is 50 W. As shown in Fig. 1, the output laser beam has a diameter of ~5 mm, collimated by a set of reflectors. To match the fiber mode field, the laser beam was focused by a plano-convex lens (L1) with a focal length of 7.5 cm, giving rise to an in-fiber coupling efficiency of ~85%. In the experiment, ~1-cm-long coating area of the AR-HCF was stripped off, so as to avoid heat-related damage due to uncoupled laser power. The bare fiber end was then mounted on a copper plate using a 3-dimensional stage (Elliot MDE122). Note that in the experiment, a water-cooling system was used to reduce thermal effects of the copper plate at high laser powers. The overall transmission efficiency of the 3-m-long AR-HCF system, including the in-fiber coupling, was measured to be ~76%. The fiber loss of the AR-HCF used in the experiment was measured to be ~0.1 dB/m at the pump wavelength of 1064 nm, see Supplement 1 for details.

The AR-HCF sample is encased within a protective metallic cable, suspended from hooks anchored to the ceiling, to ensure its vertical ingress into the micromachining unit. The minimum bending radius

As we gradually increased the input average power into the AR-HCF from 0.7 W to 20 W, we observed no significant change in both pulse temporal width and spectral profile at output port of the 3-m-long AR-HCF, thanks to the low nonlinearity and dispersion values of the fiber. In the experiment, a free-space laser beam delivery system was also constructed, and the power and pointing stability of the laser beam after transmission in the free-space system, the 3-m-long AR-HCF1 sample and the 3-m-long AR-HCF2 sample were measured for comparison. As illustrated in Fig. 2, the output light in these three cases exhibit similar power stability at a power level of ~20 W, giving root mean square (RMS) values of power fluctuations all at ~0.2% over 30-minute recording, see Figs. 2(a), 2(c), and 2(e). Note that we observed no damage of AR-HCF end-faces over several weeks of experiments and the measured RMS values remained to be almost unchanged, highlighting good power stability of the ultrafast laser fiber-delivery system.

Before used in the micromachining experiment, the laser beam profiles output from the three delivery systems were measured using a CCD camera, and their pointing stabilities (the stabilities of the laser-beam weight centers) were calculated over 30-minute recording. The results are illustrated in Figs. 2(b), 2(d) and 2(f). It can be found that the pointing stability of the free-space delivery system ($\theta_x$ = 26.3 μrad and $\theta_y$ = 17.2 μrad) is obviously worse than these of the two AR-HCF delivery systems, which can be mainly attributed to the 3-m propagation length of the laser beam in free-space. In contrast, in the fiber delivery system the output laser beam was directly coupled into the AR-HCF sample after merely a short length of free-space propagation, largely improving the pointing stability of the transmitted beam. It can be found in Fig. 2 that the performance ($\theta_x$ = 9.8 μrad and $\theta_y$ = 8.5 μrad) of the AR-HCF2 is a few better than that ($\theta_x$ = 14.0 μrad and $\theta_y$ = 9.9 μrad) of the AR-HCF1. This is because that the high-order-mode suppression ability of the AR-HCF2 (d/D = ~0.68 [19]) ensures highly-pure fundamental optical mode at the output port of the fiber.

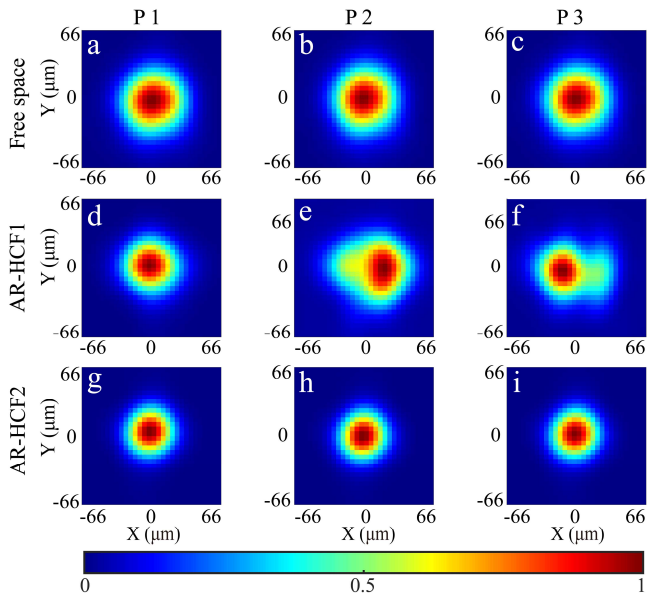

Fig. 3. Measured beam profiles at output ports of the laser delivery systems over free space (a-c), 3-m-long AR-HCF1 (d-f) and 3-m-long AR-HCF2 (g-i).

The highly-pure mode performance of the AR-HCF2 sample were further verified through measuring directly the laser beam profile at the fiber output port, as illustrated in Fig. 3. It can be found that for the free-space delivery system, the Gaussian-like beam profile remained stable (see Figs. 3(a)-3(c)) as the laser micromachining unit moved, while for the AR-HCF1 system obvious optical-mode degradation was observed (see Figs. 3(d)-3(f)) as the unit moved, which can result from some micro-bending of the AR-HCF1 sample. Such micro-bending gives rise to certain energy coupling from fundamental optical mode to some high-order modes of the fiber, leading to beam-profile instability. This beam-profile instability can be efficiently eliminated through using AR-HCF2 (see Figs. 3(g)-3(i)), since the high-order modes in this fiber has much higher loss than that of the fundamental optical mode [19,20], giving enhanced capability of high-order-mode suppression. Highly-pure fundamental optical mode with a Gaussian-like shape was always obtained at the fiber output port, immune to external perturbations.

The good laser-delivery performance of the AR-HCF2 sample was confirmed in practical laser micromachining experiments. When 2-mm-thick aluminum sheets were used as the materials under processing, we first set the laser to operate at the single-pulse model and the pulse energies on the materials were adjusted to be 60 μJ and 80 μJ. Three pieces of aluminum sheets were placed at three different positions (P1, P2 and P3) on the working platform with 16 cm intervals between them, see Fig. 1. We found in the experiment that as we move the micromachining unit to different positions, the output beam profiles from AR-HCF1 sample are different due to different micro-bending conditions of the fiber, which can lead to inconsistent single-shot processing results on the aluminum surfaces, see Figs. 4(d)-4(f). For the free-space and AR-HCF2 cases, the stable beam profiles (see Fig. 3) ensure better consistency of the single-shot processing results at both 60-μJ and 80-μJ pulse energies, see Figs. 4(a)-4(c) and Figs. 4(g)-4(i).

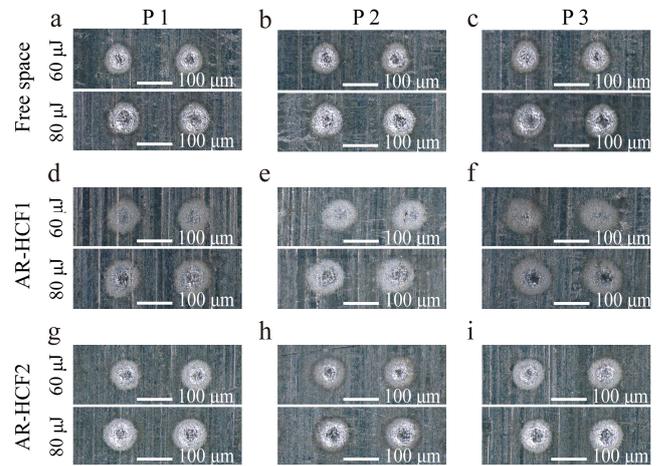

Fig. 4. Single-shot processing results on the surfaces of aluminum sheets using picosecond laser beams delivered over free space (a-c), 3-m-long AR-HCF1 (d-f) and 3-m-long AR-HCF2 (g-i).

In the experiment, we also performed the in-line processing procedure when the AR-HCF is in an oscillating state (see Fig. 1). The laser pulse-repetition-rate was set to be 1.5 MHz and the pulse energy on the aluminum surface was adjusted to be ~13 μJ, corresponding to an average laser power of ~20 W. At a scanning speed of 4 mm/s, the in-line processing results of laser beams from the free-space, 3-m-long AR-HCF1 and 3-m-long AR-HCF2 delivery systems, are illustrated in Figs. 5(a), 5(b) and 5(c) respectively. It can be seen in Fig. 5 (b) that the in-line processing results using AR-HCF1 exhibit obvious displacements from the perfectly-straight line, which can be attributed to the instability of the output beam from the fiber. As illustrated in Fig. 5(c), such displacements can be reduced through using AR-HCF2, largely improving the manufacturing quality. It can also be found from Figs. 5(a) and 5(c) that the manufacturing quality of the AR-HCF2 delivery system is almost the same with that of the free-space delivery system, with however improved configuration flexibility and reduced maintenance cost. The quality difference in the processing results of these three laser delivery systems mainly depends on the stability of the output beam. Some more detailed results and discussions are shown in Supplement 1. To quantitatively evaluate the quality of the in-line processing, we estimated the degree of displacement (defined as the ratio of the standard deviation to its mean value) for these in-line processing results, and the estimated degrees of displacement are 2.12%, 5.66%, and 2.38% for the free-space, AR-HCF1 and AR-HCF2 delivery systems, respectively. Some detailed information on the estimation of displacement degree can be found in Supplement 1.

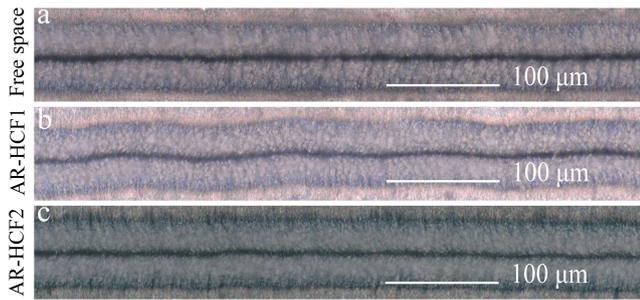

Fig. 5. In-line processing results on the surfaces of aluminum sheets using ~20 W picosecond laser beams delivered over free space (a), 3-m-long AR-HCF1 (b) and 3-m-long AR-HCF2 (c), all at 4 mm/s scanning rates.

In conclusion, we demonstrated the high-quality delivery of high-power, picosecond laser pulses over a 3-m-long AR-HCF sample for laser micromachining applications. The experiment results show that the AR-HCF sample with a d/D value of ~0.68 exhibits high optical mode purity, which is critical for overcoming the laser beam instability due to fiber micro-bending. Single-shot and in-line processing procedures of aluminum sheets were performed in our experiments, verifying that the ultrafast laser beam delivered through 3-m-long AR-HCF sample can be used to achieve similar manufacturing quality with that delivered over free space. Our findings provide some insightful guidelines for developing ultrafast laser micromachining equipment based on flexible AR-HCF delivery technique.

**Funding.** National Natural Science Foundation of China (62205353, 62275254, 61925507 and 12388102), the Strategic Priority Research Program of the Chinese Academy of Science (XDB0650000), the Shanghai Science and Technology Plan Project Funding (23JC1410100), the Fuyang High-level Talent Group Project.

# Flexible delivery of high-power picosecond laser in purely-single optical mode of anti-resonant hollow-core fiber for micromachining

XINSHUO CHANG,[1,2,3,4,†] QINAN JIANG,[1,†] ZHIYUAN HUANG,[1,2,*] JINYU PAN,[2] QINGWEI ZHANG,[5,6] NAN LI,[5] ZHUOZHAO LUO,[1] RUOCHEN YIN,[1,3] WENBIN HE,[1] JIAPENG HUANG,[1] YUXIN LENG,[2,4] XIN JIANG,[1] SHANGLU YANG,[5] AND MENG PANG[1,2,7]

[1]Russell Centre for Advanced Lightwave Science, Shanghai Institute of Optics and Fine Mechanics (SIOM) and Hangzhou Institute of Optics and Fine Mechanics (HIOM), Hangzhou 311421, China
[2]State Key Laboratory of High Field Laser Physics and CAS Center for Excellence in Ultra-intense Laser Science, Shanghai Institute of Optics and Fine Mechanics (SIOM), Chinese Academy of Sciences (CAS), Shanghai 201800, China
[3]Center of Materials Science and Optoelectronics Engineering, University of Chinese Academy of Sciences, Beijing 100049, China
[4]Hangzhou Institute for Advanced Study, Chinese Academy of Sciences, Hangzhou 310024, China
[5]Center for Laser-Aided Manufacturing and Processing, Shanghai Institute of Optics and Fine Mechanics (SIOM), Chinese Academy of Sciences (CAS), Shanghai 201800, China
[6]e-mail: zhangqingwei@siom.ac.cn
[7]e-mail: pangmeng@siom.ac.cn
*Corresponding author: huangzhiyuan@siom.ac.cn
†These authors contributed equally to this work.

# Flexible delivery of high-power picosecond laser in purely-single optical mode of anti-resonant hollow-core fiber for micromachining: supplement document

## 1. The measured loss of the anti-resonant hollow-core fiber (AR-HCF)

The measured loss of the AR-HCF was obtained using the cut-back method, and the measurement results were plotted in Fig. S1. Figures S1(a) and S1(b) correspond to the fiber losses of AR-HCF1 and AR-HCF2, respectively. The dashed black lines in Figs. S1(a) and S1(b) indicate the pump wavelength of 1064 nm, at which the fiber losses of AR-HCF1 and AR-HCF2 were measured to be 0.12 dB/m and 0.10 dB/m, respectively.

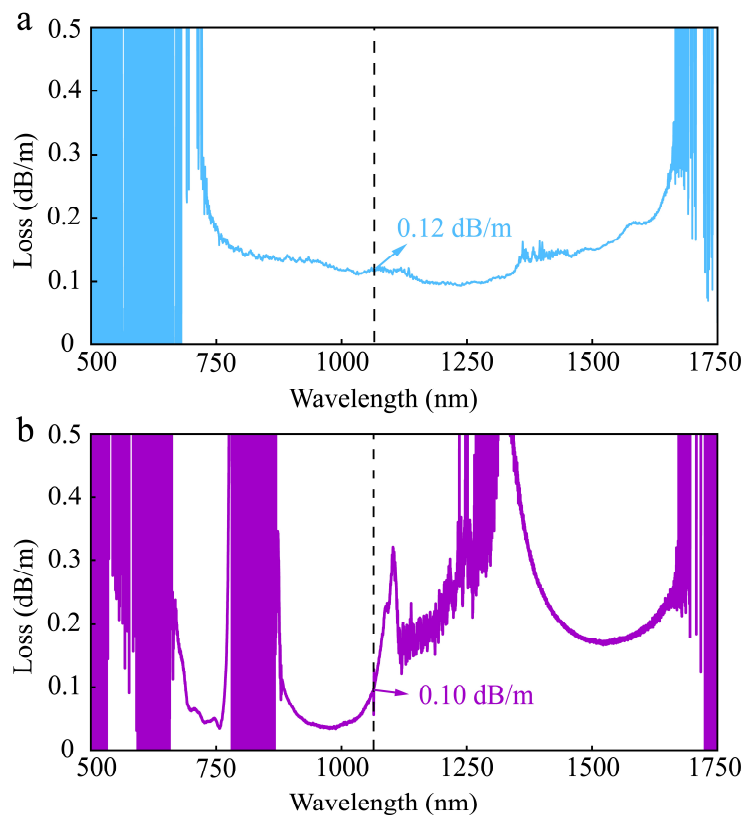

**Fig. S1.** Measured loss of the AR-HCF, (a) and (b) correspond to AR-HCF1 and AR-HCF2, respectively. The dashed black lines in panels (a) and (b) represent the pump wavelength of 1064 nm, at which the fiber losses of AR-HCF1 and AR-HCF2 were measured to be 0.12 dB/m and 0.10 dB/m, respectively.

## 2. Evolution of beam profiles at output ports of three laser delivery systems in static and oscillating states

Figures S2(a) and S2(b) present the evolution of beam profiles measured at 20-second intervals from the output ports of three laser delivery systems (including free space, 3-m-long AR-HCF1, and 3-m-long AR-HCF2) under static and oscillating fiber conditions, respectively. When the AR-HCF is in a static state, it can be observed that for the laser delivery systems: 3-m-long AR-HCF1 and 3-m-long AR-HCF2, the beam profiles at the output ports during in-line processing are as stable as the beam profiles in free space, as shown in Fig. S2(a). However, when the AR-HCF is in an oscillating state, we can observe the beam-profile instability at the output port of the AR-HCF1, which is caused by certain energy coupling from fundamental optical mode to some high-order modes of the fiber (see Fig. S2(b)). Therefore, the in-line processing results using AR-HCF1 exhibit obvious displacements from the perfectly-straight line due to the instability of the output beam from the fiber, see Fig. 5(b) of the main text. While for AR-HCF2, this beam-profile instability can be efficiently eliminated, so that highly-pure fundamental optical mode with a Gaussian-like shape can always be obtained at the fiber output port without being affected by external perturbations, see Fig. S2(b). It can be found that the manufacturing quality of the AR-HCF2 delivery system is almost the same with that of the free-space delivery system, see Figs. 5(a) and 5(c) of the main text.

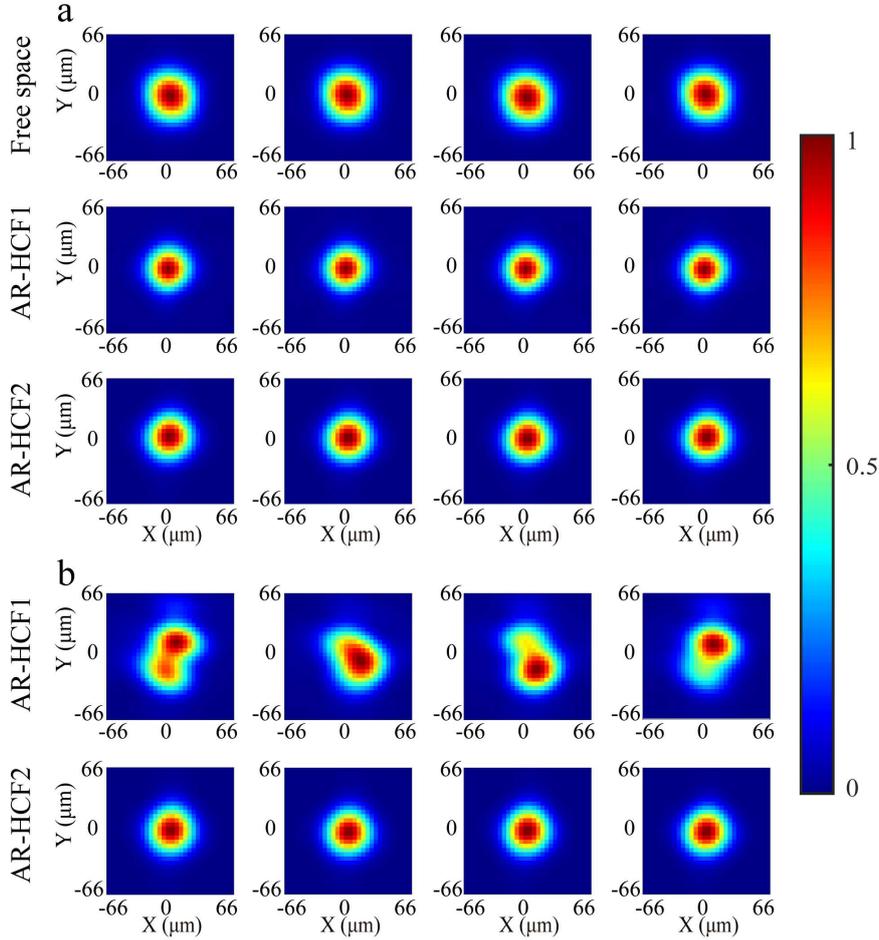

**Fig. S2.** During in-line processing, beam profiles were measured at 20-second intervals from the output ports of three laser delivery systems: free space, 3-m-long AR-HCF1, and 3-m-long AR-HCF2, under static (a) and oscillating (b) fiber conditions. The measured beam profiles in (b) correspond to the in-line processing results in Figs. 5(b) and 5(c) of the main text.

## 3. Estimation of displacement degree

In order to quantitatively evaluate the quality of the in-line processing, we estimated the displacement degree of these in-line processing results. Figure S3 exhibits the in-line processing result on the surfaces of aluminum sheet using ~20 W picosecond laser beams delivered over 3-m-long AR-HCF2. We first select two points at the beginning and end of this processing trace and connect them into a line, using this line as the reference line (solid red line in Fig. S3). Then, 52 points are uniformly selected along the center of this processing trace, and the vertical distances from these points to the reference line are obtained (the vertical distances of 8 sampling points are shown in Figs. S3). The displacement degree of this processing result is defined as the ratio of the standard deviation of these 52 values to their mean, and the calculated value is 2.38%. Using the same estimation method, we can obtain the displacement degrees are 2.12% and 5.66% for the laser delivery systems over free space and 3-m-long AR-HCF1, respectively.

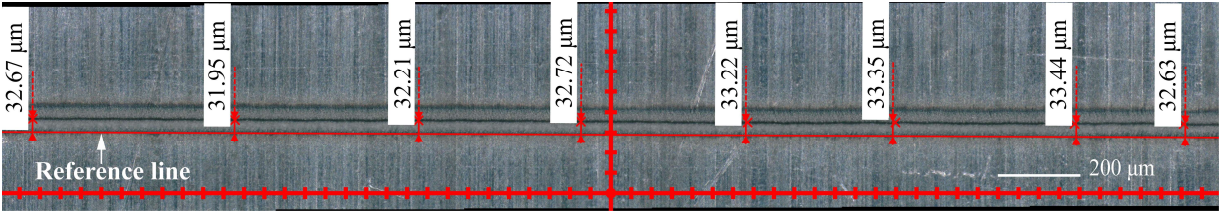

**Fig. S3.** Schematic diagram of point-based estimation of displacement degree. This in-line processing result on the surfaces of aluminum sheet using ~20 W picosecond laser beams delivered over 3-m-long AR-HCF2.